# Joint Resource Bidding and Tipping Strategies in Multi-hop Cognitive Networks

Beatriz Lorenzo, Ivana Kovacevic, Ana Peleteiro, Francisco J. González-Castaño, Juan C. Burguillo

*Abstract*—In multi-hop secondary networks, bidding strategies for spectrum auction, route selection and relaying incentives should be jointly considered to establish multi-hop communication. In this paper, a framework for joint resource bidding and tipping is developed where users iteratively revise their strategies, which include bidding and incentivizing relays, to achieve their Quality of Service (QoS) requirements. A bidding language is designed to generalize secondary users' heterogeneous demands for multiple resources and willingness to pay. Then, group partitioning-based auction mechanisms are presented to exploit the heterogeneity of SU demands in multi-hop secondary networks. These mechanisms include primary operator (PO) strategies based on static and dynamic partition schemes combined with new payment mechanisms to obtain high revenue and fairly allocate the resources. The proposed auction schemes stimulate the participation of SUs and provide high revenue for the PO while maximizing the social welfare. Besides, they satisfy the properties of truthfulness, individual rationality and computational tractability. Simulation results have shown that for highly demanding users the static group scheme achieves 150% more winners and 3 times higher revenue for the PO compared to a scheme without grouping. For lowly demanding users, the PO may keep similar revenue with the dynamic scheme by lowering 50% the price per channel as the number of winners will increase proportionally.

## I. INTRODUCTION

New economic models and system architectures have emerged to better manage spectrum resources [1], [2]. Among these proposals, auction mechanisms have attracted much attention as an efficient approach to pricing and resource allocation. In spectrum trading markets, a spectrum owner or primary operator (PO) leases its idle licensed spectrum bands to secondary users (SUs) to obtain profit [3]. As leased spectrum usage is fundamentally opportunistic, the SU must assess its needs, determine the level of uncertainty that it can tolerate and decide whether the spectrum quality is worth its price.

Auction in single-hop networks where SUs request a single channel has been covered extensively in the literature and is well understood. However with the advent of new communication paradigms, an SU may have multi-hop access to base stations, access points or other users [4], [5]. The main challenge for spectrum trading in multi-hop cognitive networks is to establish multi-hop communication using the purchased channels under uncertain availability of licensed spectrum. Besides, primary user (PU) return has significant impact on the opportunistic usage of licensed spectrum and the achievable quality of service (QoS). Therefore, SUs may need more than one channel as backup to mitigate these effects and keep delay at reasonable levels. The importance of backup channels to increase link robustness in cognitive networks is addressed in [6]. The impact of the number of channels on sale is studied in [7] by considering a three-layer spectrum market consisting of the spectrum holders, service provider and end users. However, most works on spectrum auction assume that one buyer can bid for at most one channel [8] while others [9], [10] assume that a buyer can place bids for multiple channels but win only one. Few recent works consider combinatorial auctions [11] in which buyers may win more than one channel, but incur heavy computational overheads. The limitations of the previous schemes make them impractical for multi-hop networks with buyers bidding for several channels.

The few works that do consider spectrum auction in multi-hop networks [12], [13] ignored both the uncertainty of channel availability due to PU activity and QoS provisioning. They simply focused on providing the best channel allocation and pricing according to interference. Besides, the complexity of these schemes significantly grows with the number of bidders, making them unsuitable for large networks.

In this paper, we aim to address the above issues. Specifically, we study spectrum auctions in multi-hop cognitive cellular networks where SUs have different QoS requirements in terms of delay and thus, heterogeneous valuations and willingness to pay. A framework for joint bidding and tipping is developed to encourage users to iteratively revise their strategies, which include bidding and incentivizing relays, to achieve their QoS requirements. A bidding language is designed to generalize heterogeneous SU demands for multiple resources. In order to provide high revenue to the PO and exploit users' heterogeneity, static and dynamic group partition schemes are developed together with new winner selection and payment mechanisms. The concept of group buying emerged from Internet services such as Groupon [14]. Few works have considered it for spectrum allocation [15], [16]. The purpose of these works is to apply group buying exclusively to price reduction. However, in addition to the above advantage, in our schemes the PO performs the grouping and determines the group partition strategy based on users' QoS requirements to fairly allocate resources and increase its own revenue. We prove that the auction schemes not only provide high revenue for the operator but also maximize social welfare and satisfy the properties of individual rationality, truthfulness and computational efficiency. This last property is highly desirable for auctions in multi-hop networks.

The main contributions of the paper can be summarized as follows.

1) A network model for multi-hop cognitive cellular networks (MC$^2$Ns) that enables analysis of a number of performance metrics as a function of the number of purchased channels and users´ availability to relay. This model facilitates a tractable analysis of the network and provides insights into the resources needed to auction based on users´ QoS requirements.

2) A joint bidding and tipping scheme that relies on a new bidding language where SUs bid for multiple commodities and incentivize users to relay data to achieve their QoS requirements.

3) New group partitioning-based auction mechanisms for multi-hop networks. These mechanisms consider static and dynamic partition strategies for spectrum allocation to exploit the heterogeneous QoS demands of SUs. The groups are formed on the basis of resource reusability. In the static scheme, the winning groups are chosen to preserve the QoS requirements of the highest bidders and the dynamic scheme is intended for users with low QoS requirements and a tight budget. Hence, the proposed schemes achieve different tradeoffs between revenue and fairness.

4) Development of a reinforcement learning algorithm to automate the bidding process based on users' previous experiences and compare it to the iterative joint bidding and tipping scheme.

Extensive simulations have been carried out to show the performance of our schemes and highlight their robustness in large networks compared with existing schemes.

The rest of this paper is organized as follows. The multi-hop model is described in Section II. In Section III, the spectrum-aware routing discovery protocol is characterized. The joint bidding and tipping scheme is developed in Section IV. In Section V, the group partitioning-based auction schemes are presented. Numerical results are given in Section VI. In Section VII we discuss related work. Finally, Section VIII concludes the paper.

## II. NETWORK MODEL

### A. Secondary Spectrum Market

We consider a multi-hop cognitive cellular network (MC$^2$N) as presented in Fig. 1, where users are uniformly distributed across the cell. For modeling purposes, the area of the cell is divided into hexagonal subcells of radius $r$. In each subcell, there is, potentially, an SU that will act as a source, relay or destination. We remark that this model is intended to support the evaluation of the auction schemes presented and its complexity is deliberately reduced to offer valuable insights into the performance of these schemes without entering into unnecessary details.

Each SU is equipped with one radio capable of switching between different channels. We assume that a spectrum owner or PO leases its idle channels to SU sources. Due to PU activity, there is uncertainty regarding channel availability (PU activity is modeled in Section II.B). The SU *source* will transmit uplink by relaying to one of its adjacent SUs (located in adjacent subcells) by using the available channels (one channel per hop). In the adjacent subcell, there will be a SU available to relay with probability $p$, which depends on coverage, battery charge, and willingness to cooperate.

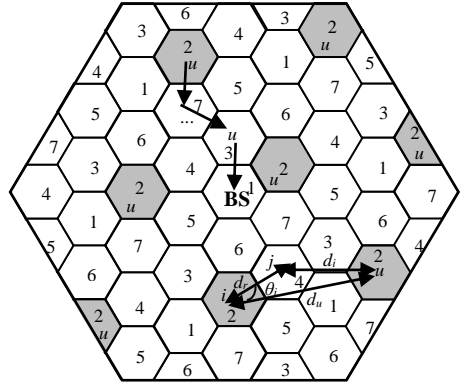

Fig. 1. Network model with $K$-reuse pattern for $K = 7$.

We assume that SUs have QoS requirements in terms of delay. To mitigate the impact of PU return, SUs keep a number of frequency channels as backup. Spectrum trading in a multi-hop secondary network involves decisions by SU sources on the number of channels to bid for, route selection and required relay availability probability $p$. In the bidding process, each source node submits a *compound bid* $B = \{\gamma, R, bid\}$ to the auctioneer (PO) for a number of resources $\gamma = \{b, \tau_{max}, P\}$ needed to transmit through the preferred route $R$ for a maximum time interval $\tau_{max}$ (QoS requirement), where $b$ denotes the number of channels and $P$ the transmission power. According to the impossibility theorem [17] we cannot simultaneously maximize social welfare and operator revenue. Since our focus is on designing auction mechanisms for multi-hop secondary networks, we aim to maximize social welfare. This is a common assumption in the literature when designing auction mechanisms for secondary networks [12], [13]. We assume that the PO leases its idle channels without causing quality degradation to its own services and thus, the PO's revenue is the total payment received from the winning SUs. Nevertheless, we are aware that the PO will be interested in maximizing its revenue. Consequently, we develop partition schemes to provide high revenue to the PO by exploiting the heterogeneity of SUs' demands and willingness to pay. After the auctioneer collects all the bids, determines the *partition strategy* (i.e., static or dynamic) that provides the highest revenue. Finally, the auctioneer assigns the resources to the winning SUs to maximize social welfare.

Winning sources will pay a *price* to the auctioneer and transmit the traffic to the adjacent user on the route by utilizing the purchased resources. In addition, they will pay some incentive *tip* to the relays to encourage them to participate in the transmission. Consequently, the SUs' strategy will consist of determining the minimum bid and tip to satisfy their QoS requirements. Let us remark that the implementation of the multi-hop transmissions does not depend on the partition strategy used.

### B. Cognitive Link Availability

We model the activity of PUs since the transmission of SUs on cognitive links depends on channel availability. We assume that the PO will lease channels not occupied by any PU to avoid degrading the performance of its own licensed users. This is a common approach to modeling the interaction between the PO and the SUs as a trading process [12], [13].

To model the link availability in a network with multiple channels $c$, we adopt an M/M/c queuing model as in [6], [18]. By modelling arrivals as a Poisson process and call/session duration as exponentially distributed, the probability that the SU will have $b$ channels available, $p_b$, can be obtained as a solution of birth/death equations [6]. Note that the schemes presented in this paper are valid for simpler traffic models for PU activity, i.e., ON/OFF model [5].

If we assume that spectrum monitoring is perfect, the probability that a PU will return to a channel currently allocated to an SU is denoted by $p_{return}$. This is checked at every hop of the route and is obtained as in [6]. If a PU returns to the channel currently allocated to an SU, it will interrupt the transmission and force the SU to try a new option. PU activity will have an impact on the number of channels needed by SU sources to satisfy their QoS requirements and is considered in the route discovery protocol in the following section. For the tractability of the problem, we consider that the probability of PU return on a channel is the same for every channel and thus, all channels are identically desirable.

### C. Tessellation Factor and Scheduling

We assume users are interested in transmitting with the minimum power possible $P_i = P_{i,min}$ to limit interference and power consumption. For simplicity, we assume that all users transmit with the same power $P_i = P$.

The Shannon capacity on link $l$ when transmitting on channel $\psi$ is given by $c_l = \log(1 + SINR(\psi, P))$ where

$$SINR(\psi, P) = \frac{PG_{ij}^{\psi}}{\sum_{u=1}^{n_\psi} PG_{uj}^{\psi} + \chi_r}$$

is the signal to interference noise ratio at any relaying user, $G_{ij}^{\psi}$ is the channel gain between user $i$ and $j$ on channel $\psi$, $n_\psi$ is the number of concurrent transmissions using that channel, $G_{uj}^{\psi}$ is the channel gain between interfering users $u$ and $j$ on channel $\psi$ and $\chi_r$ is the background noise power. The detailed calculus of the interference given the geometry of the cell is provided in [6]. Then, the capacity on the multi-hop route $R$ is $c_R = \min_{l \in R} c_l$.

The efficiency of capacity usage when $b$ channels are used on the route can be obtained as

$$\bar{c}_R = c_R / b. \quad (1)$$

The optimization of scheduling in a multi-channel multi-hop network is an NP-hard problem [5]. To keep the scheduling process simple, we applied a conventional resource reuse scheme for cellular networks to our tessellation model, shown in Fig. 1, for a resource reuse factor $K = 7$. The subcell index within the cluster indicates the slot allocation, $k = 1,..., K$. Users with the same slot index will transmit simultaneously. The transmission turn (in a round robin fashion) is given by the slot index. When a user is scheduled for transmission, it will transmit its own traffic and the relayed traffic. To avoid transmission/reception collisions, users can transmit simultaneously if they are separated by a distance $d > 2d_r$, where $d_r$ is the relaying distance between adjacent subcells, $d_r = \sqrt{3} \cdot r$ [4] and $r$ is the subcell radius. This constraint is a direct consequence of the fact that users are equipped with a single radio. In this way, transmissions will be collision-free regardless of the adjacent relay or the channel assigned for the transmission. This constraint holds for $K \geq 7$. To take advantage of all available channels and reduce interference in the network, the PO will randomly assign a different channel to each subcell where there is a winning SU sharing transmission slot. The $K$-scheduling pattern proposed is a heuristic that avoids the complexity of re-computing the schedule according to PU activity and user availability to relay.

The terminology used in the paper is summarized in Table I.

Table I. Notation

| Parameter | Definition |
|---|---|
| $r$ | Subcell radius |
| $K$ | Reuse pattern |
| $P$ | Transmission power |
| $c_R, \bar{c}_R$ | Route capacity, efficiency of route capacity usage |
| $N$ | Number of subcells |
| $b$ | Available channels |
| $p$ | Relay availability probability |
| $p_{return}$ | Probability of PU return to a channel currently allocated to an SU |
| $p_{free}$ | Probability of free channel |
| $\tau, \tau_{max}$ | Delay, maximum tolerable delay |
| $\xi_m, \tilde{\xi}_m$ | Strategy of source $m$, winning strategy |
| $\beta_m, \tilde{\beta}_m$ | Bidding strategy of source $m$, winning bidding strategy |
| $\theta_m, \tilde{\theta}_m$ | Tipping strategy of source $m$, winning tipping strategy |
| $\bar{C}_e$ | Normalized effective route capacity |
| $p_{D,m}$ | Probability that $m$ access the destination |
| $\Delta_m$ | Scheduling delay |
| $V_m$ | Valuation function of source $m$ |
| $U_m$ | Utility function of source $m$ |
| $price_c$ | Tentative clearing price |
| $price_q$ | Clearing price |
| $\delta$ | Price step |
| $\mathcal{N}_k$ | Set of SUs sources in partition $k$ |
| $\Theta_k$ | Bid of group $k$ |
| $S$ | Number of winning groups |

### III. SPECTRUM-AWARE ROUTING DISCOVERY PROTOCOL

This section presents and analyzes a route discovery protocol for SUs. By following the $K$-scheduling pattern, collisions between adjacent relays are avoided and thus, we focus on discovering the route based on users' availability to relay and PU activity. The analysis of the protocol provides the delay and the probability of accessing the destination, both crucial metrics for estimating the needed resources to be auctioned.

## A. Description of the Protocol

We assume that the PO shares with potential SUs information regarding the number of channels $b$ available for auction and the channel uncertainty given by $p_{return}$. This way the SU is released from sensing and any errors it may experience.

Let us assume that in subcell $i$ there is an SU willing to transmit to an intended destination by relaying to adjacent users. As each user has 6 adjacent subcells, the candidate relay may be in any of these subcells $w$, $w = 1,.., 6$. We start by assuming that the route discovery protocol provides the shortest available path, although the protocol admits other criteria for relay selection. Then, according to the relay priorities given by the distance to the destination, the SU first checks the availability of the adjacent user (located in the adjacent subcell) in the direction corresponding to *the shortest distance* towards the destination. The user will be available to relay with probability $p$. If this is the case, the SU will transmit to the adjacent subcell through a cognitive channel, and this transition will occur with probability $p_{w=1}$. If the user is not available, the SU will check the availability of the next adjacent user (2nd user), as shown in Fig. 2. If that user is available, the SU will relay to it. This transition will occur with probability $p_{w=2}$. Otherwise, the protocol will continue in the same fashion until it checks the last adjacent user (6th user). If in the last adjacent subcell there is no user available to relay, the route will not be established, which happens with probability $p_0$. The same process is repeated at every hop of the route.

As channel availability depends on the activity of the PUs, in any system state a PU may return to the channel after the transmission to a relay is initiated. This is also illustrated in Fig. 2. In this case, the process will be aborted (with probability $p_{return}$) and the SU will try another channel.

The relaying probability from subcell $i$ to any adjacent subcell $j$ can be obtained by mapping the transmission pair $(i,j) \rightarrow w$, as

$$p_w(b) = p_b p(1-p)^{w-1} p_{free}(b), \ w=1,...,6 \quad (2)$$

where $p_b$ is the probability that the SU has $b$ channels available given in [6], $p$ is the probability that in the adjacent subcell there is a user willing to relay, and $p_{free}$ is the probability that there will not be a PU return to the channel, $p_{free} = 1 - p_{return}$, and $p_{return}$ is obtained as in [6].

The probability that the SU will not be able to transmit to any adjacent user is $p_0(b) = 1 - \sum_w p_w(b)$, which is represented as a transition to an absorbing state $nr$ (no route).

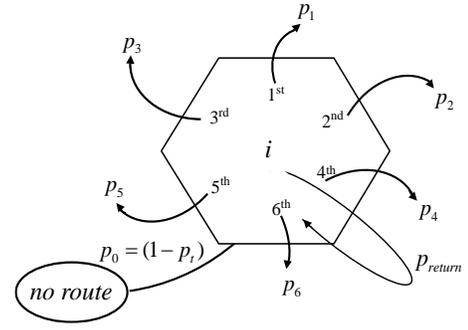

Fig. 2. State transition probabilities of the Markov model as in (2).

## B. Analysis of the Protocol

We define a relaying probability matrix $\mathbf{P}(b) = [p_{ij}(b)]$ where entry $p_{ij}(b)$ indicates the relaying probability from subcell $i$ to $j$ when $b$ channels are available. This probability can be obtained from (2) by mapping the transmission from subcell $i$ to any adjacent subcell $j$ as $(i,j) \rightarrow w$. It is worth noticing that $\mathbf{P}$ also defines the network topology. To analyze the relaying process in the network, we map the tessellation scheme into an absorbing Markov chain with a set of absorbing states $\mathcal{A}_s = \{D, nr\}$. These states represent the end of the route when the user has reached the destination (e.g., BS, AP, mobile to mobile connection) or when no route ($nr$) is available.

Then, we reorganize the relaying matrix into an $(N + 1) \times (N + 1)$ matrix of the form [19]

$$\mathbf{P}^*(b) = \begin{bmatrix} \mathbf{I} & \mathbf{0} \\ \mathbf{R}(b) & \mathbf{Q}(b) \end{bmatrix} \quad (3)$$

where $N$ is the number of subcells, $\mathbf{I}$ is an $N_A \times N_A$ diagonal unitary matrix corresponding to the number of absorbing states, $\mathbf{0}$ is an $N_A \times (N - N_A + 1)$ all-zero matrix, $\mathbf{R}$ is the $(N - N_A + 1) \times N_A$ matrix of transition probabilities from transient states to absorbing states and $\mathbf{Q}$ is the $(N - N_A + 1) \times (N - N_A + 1)$ matrix of transition probabilities between transient states.

By defining $\mathbf{N} = (\mathbf{I} - \mathbf{Q})^{-1}$, the mean time for the process to reach any absorbing state starting from transient state $i$ is [19]

$$\boldsymbol{\tau}(b) = (\tau_1(b),...,\tau_{N-N_A+1}(b))^t = T(\mathbf{I} - \mathbf{Q}(b))^{-1}\mathbf{1} = T\mathbf{N}(b)\mathbf{1} \quad (4)$$

when the dwell time for any state $i$ is the same, $T = T_i$ and $\mathbf{1}$ is an $(N - N_A + 1) \times 1$ column vector of ones. Otherwise, $\boldsymbol{\tau} = (\mathbf{I} - \mathbf{Q})^{-1}\boldsymbol{v} = \mathbf{N}\boldsymbol{v}$ where $\boldsymbol{v}$ is a column vector whose components are $T_i$.

For the normalized dwell time $T = T_i = 1$, the entries $\tau_i$ of vector $\boldsymbol{\tau}$ represent the average number of hops from state $i$ to the absorbing state. This expression will be used in the next section to define SU valuation.

The probability that the Markov process starting in a transient state $i$ ends up in an absorbing state $j$ is $e_{ij}$, where

$$\mathbf{E} = [e_{ij}] = (\mathbf{I} - \mathbf{Q})^{-1}\mathbf{R} \quad (5)$$

The probabilities of accessing the destination and no route availability are

$$[\mathbf{p}_D, \mathbf{p}_{nr}] = \mathbf{f}\mathbf{E} \quad (6)$$

where $\mathbf{f}$ is the vector of probabilities of initial user positions, $\mathbf{p}_D = (p_{D,1},...,p_{D,N-N_A+1})^t$ and $\mathbf{p}_{nr} = (p_{nr,1},...,p_{nr,N-N_A+1})^t$.

## C. Multi Session Routing

When multiple routes are simultaneously active in the network there is a probability that two routes will favor simultaneously a given subcell while the relay can only be available for one route. The remaining routes will look for another subcell in accordance with our route discovery protocol. This phenomenon can be modelled by modifying the relay availability probability $p$ as follows. A subcell used by route $m_1$ will be needed also by route $m_2$ with $l_{m2} - 1$ relays with probability $(l_{m2} - 1)/N$ where $l_{m2}$ is the route length. If $M$ is the number of routes simultaneously active in the network, the subcell required by route $m_1$ will not be required by any other route with probability $\prod_{m_2=1, m_2 \neq m_1}^{M} (1-(l_{m_2}-1)/N)$. Therefore, if $M$ routes are simultaneously active in a network the relay availability probability should be modified as $p \to p \prod_{m_2=1, m_2 \neq m_1}^{M} (1-(l_{m_2}-1)/N)$. By using the new $p$ in (2) the same analysis applies. The delay and probability of accessing the destination are obtained by (4) and (6), respectively.

By using alternative routes provided by the route discovery protocol we can avoid potential route collisions. In addition, in a high dense cognitive network we can exploit the frequency channels available per subcell by letting different users per subcell relay on a different frequency channel simultaneously or considering more advanced SU terminals equipped with multiple antennas. If there are $b$ channels available in the network then route $m_2$ will not be able to use a specific subcell only if all $b$ channels are allocated to other routes.

The probability of having $b$ or more routes attempting to use the same subcell is

$$p_c = \sum_{i=b}^{M} \binom{M}{i} \left(\frac{l_{m2}-1}{N}\right)^i \left(1 - \frac{l_{m2}-1}{N}\right)^{M-i}$$

The availability probability $p$ can now be modeled as $p \to p(1-p_c)$. The available channels would be used as backup to avoid collisions as well as PU returns. Even so, if a collision occurs, the route will not be established with probability $p_0$ as described in Section III.A. For clarity of presentation, we have not elaborated this case but the extension in this direction is straightforward.

## IV. JOINT BIDDING AND TIPPING SCHEME

In this section, we first describe the desired economic properties of our auction schemes and review some concepts, and then present the joint bidding and tipping scheme. The auction is a sealed-bid auction with one auctioneer (PO) and multiple bidders (SU sources) requesting multiple channels. The auction procedure consists of a winner selection process (resource allocation rule) and a payment mechanism.

### A. Design Goals

We assume that bidders are selfish and may lie about their valuation to maximize their utility. We define the dominant strategy of a player as the one that maximizes the player's utility regardless of what other players' strategies are. Formally, if $\xi_m$ is the strategy of player $m$, for any $\xi_m' \neq \xi_m$, and any strategy profile of other players $\xi_{-m}$, we have $u_m(\xi_m, \xi_{-m}) \geq u_m(\xi_m', \xi_{-m})$. If the inequality always holds, then $\xi_m$ is a strongly dominant strategy. Otherwise, it is a weakly dominant one. We aim to design auction schemes that can satisfy the following economic requirements: truthfulness, individual rationality and computational efficiency, which are defined as follows,

- *Truthfulness:* An auction is truthful if any player's true valuation is its dominant strategy. This means that given the auction rules and the strategy profiles of other players, a player cannot improve its utility by submitting a bid different from its true bid. Truthfulness is the most desirable property as it simplifies player strategies.

- *Individual rationality:* An auction is individually rational, if no bidder is charged higher than its bid.

- *Computationally efficient:* The result of the auction can be obtained in polynomial time.

Let us denote by $price_q$ the clearing price the auctioneer charges the SU source per channel. We define the PO's revenue as the total payment received from the winning SUs. Since we consider that the PO leases its own idle spectrum bands without causing quality degradation to its own services, its revenue is always non-negative. As already mentioned, due to the impossibility theorem [17] we cannot simultaneously maximize social welfare and operator revenue. Therefore, our auction schemes are designed to maximize social welfare and provide high revenue to the PO by incorporating partition schemes that exploit the heterogeneity of SU demands.

### B. Joint Bidding and Tipping Scheme

We assume that all sources bid for resources simultaneously. We define a new bidding language for multi-hop secondary networks which enables each SU source $m$ to submit a compound bid to the auctioneer as $B_m = \{\gamma_m, R_m, bid_m\}$ where $\gamma_m$ are the resources to auction, $R_m$ is the transmission route and $bid_m$ is the bid amount. In our auction schemes, the resources $\gamma_m$ are given by $\gamma_m = \{b_m, \tau_{max,m}, P_m\}$, where $b_m$ are the channels needed during maximum $\tau_{max,m}$ slots (QoS requirement) when transmitting with power $P_m = P$.

The source node determines the bid based on its valuation of the resources and willingness to pay. Besides, it also provides a *tip* to encourage the relays to cooperate and thus, reduce the delay. We define the strategy profile of source $m$ as, $\xi_m = (\beta_m, \theta_m)$ where $\beta_m$ indicates the percentage of the valuation that $m$ is willing to pay per channel and $\theta_m$ is the percentage offered as a tip per hop.

The PO is interested in the optimum allocation of resources so SUs will achieve good performance and, thus, be able to offer high bids.

The valuation function of bidder $m$ depends on the number of demanded channels $b_m$ and the relay availability probability $p$. We formulate it as

$$V_m(b_m, p) = \alpha_m \frac{\bar{C}_{e,m}(b_m, p)}{\Delta_m(b_m, p) \cdot P_{t,m}} \quad (7)$$

where,

- $\alpha_m$ is the private valuation of resources, $0 \leq \alpha_m \leq 1$, which shows the heterogeneity of users' valuations. We set $\alpha_m = 1 / \tau_{max,m}$ to model the relation between highly demanding users and high valuation.
- $\bar{C}_{e,m}$ is the effective route capacity obtained as $\bar{C}_{e,m} = p_{D,m} \cdot \bar{c}_{R_m}$, where $p_{D,m}$ is the probability of accessing the destination given by (6) and $\bar{c}_{R_m}$ the efficiency of route capacity usage[2] (1).
- $\Delta_m$ denotes the scheduling delay of route $m$ and is obtained as $\Delta_i(b_m, p) = K\tau_m(b_m, p)$ where $\tau_m$ is given by (4).
- The overall power consumption of route $m$ is $P_{t,m} = P\tau_m$.

The previous definition of valuation, in terms of throughput per unit power, has been used in multi-hop wireless networks to analyze different trade-offs [32], [33].

The bid offered is a percentage of the valuation of the resources when $b_m$ channels are used,

$$bid_m(b_m, p) = \beta_m \cdot b_m \cdot V_m(b_m, p) \quad (8)$$

where $\beta_m$ indicates the percentage of the gain that SU $m$ is willing to pay per channel. The PO will ask for a price $q$ that represents the minimum payment it is willing to accept for selling the channels. This will result in a percentage of the SU's gain, $\beta_{m,q}$.

The overall tip that source $m$ will pay to encourage users to relay with probability $p$ is

$$tip_m(b_m, p) = \theta_m \cdot \tau_m \cdot V_m(b_m, p) \quad (9)$$

where $\theta_m$ indicates the percentage of the valuation SU $m$ will offer as a tip per hop and $\tau_m$ is the number of hops on the route (4). Note that a high valuation $\alpha_m$ will result in a higher bid and tip.

Following a particular strategy $\xi_m = (\beta_m, \theta_m)$, each SU $m$ determines the number of channels $b_m$ and $p$ to satisfy its QoS requirements by solving the following optimization problem,

$$\underset{b_m, p}{\text{maximize}} \quad U_m(b_m, p) = V_m(b_m, p) - bid_m(b_m, p) - tip_m(b_m, p)$$
$$\text{subject to} \quad K \cdot \tau_m(b_m, p) \leq \tau_{max,m}$$
$$1 \leq b_m \leq c - n \quad (10)$$
$$0 < p \leq 1$$

where $\tau_{max,m}$ is the QoS constraint given in terms of delay and $c$ are the total channels in the cellular network, $n$ of which are occupied by PUs. This optimization provides the optimum number of channels $b_m$ and the optimum availability probability $p$ such that the SU obtains the maximum utility for a given value of $\beta_m$ and $\theta_m$. The valuation function defined by

---
[2] The SUs calculate an estimation of their link capacity [6] considering that in the worst case there will be $N/K$ secondary users transmitting in the same slot on any of the $b$ available channels, where $N$ is the number of subcells and $K$ the resource reuse factor. Due to the symmetry of the scenario, the same capacity is assumed per hop. All SUs estimate the capacity in the same way and thus, they bid under the same conditions.

---

(7) is concave and, thus, cost and utility functions are concave as well. Problem (10) can be solved by introducing a penalty term into the utility when the first constraint is violated. Then, the solution will favor the values of $b_m$ and $p$ such that $K \cdot \tau_m(b_m, p) \approx \tau_{max}$. The equivalent optimization problem can be solved efficiently by numerical methods [30].

*Winner Selection:* After the auctioneer receives all bids, it sorts them in decreasing order $\{bid_m\}$, $m = 1, ..., N$ and selects the $M$ highest bidders as *potential* winners. Since the $K$-reuse pattern provides full reusability of resources and eliminates collision, the auctioneer selects the potential winners based exclusively on their bids. Then, the auctioneer announces the tentative clearing price (per channel), $price_q$, and allows *all* users to *revise* their strategies and submit a new bid in the next iteration, as shown in Algorithm 1. To encourage truthfulness and obtain high revenue, the *tentative* clearing price in each iteration $t$ is increased as $price_c(t+1) = price_q(t) + \delta \cdot t$, where $\delta$ is a constant. The auction iterates until the difference in the clearing price, $|price_c(t) - price_c(t-1)| < \varepsilon$. When the auction finishes, the users whose bids are higher than the clearing price are the winners. To control the communication overhead, the PO is in charge of deciding the number of winners $M$, the step price $\delta$ and the iterations $T$ needed to satisfy its expected revenue as a function of $N$.

*Payment Mechanism:* We designed the payment mechanism to be relatively independent of winner selection to ensure truthfulness. We choose the highest losing bid as the clearing price, $price_q$. In our iterative auction, this is highest bid from the bidder who lost in all previous iterations, as given in Algorithm 1.

*Revision of strategy:* As a result of the auction in each iteration, the SU $m$ revises its strategy $\xi_m(t+1) = (\beta_m(t+1), \theta_m(t+1))$ and obtains $b^*(t+1)$ and $p^*(t+1)$ by solving (10). The strategy is updated as follows:

- Based on the tentative clearing price, $price_c(t+1)$, the new percentage of valuation offered per channel is given by (8),

$$\beta_m(t+1) \geq price_c(t+1) / b_m^*(t)V_m(t). \quad (11)$$

- Since $p^*(t)$ is the required relay availability probability, let us denote by $1 - p^*(t)$ the utilization of the relay, defined as the probability that the relay is transmitting its own traffic. The percentage of valuation offered as a tip to compensate the user for relaying is set to

$$\theta_m(t+1) = 1 - p^*(t). \quad (12)$$

After revising the strategy, the optimization in (10) is solved and the SU $m$ offers a new bid and tip, respectively, as

$$bid_m(t+1) = \beta_m(t+1) \cdot b_m^*(t+1) \cdot V_m(t+1)$$

$$tip_m(t+1) = \theta_m(t+1) \cdot \tau_m(b_m^*(t+1), p^*(t+1)) \cdot V_m(t+1)$$

If the SU cannot pay the price asked for the PO, it will not get the resources, $U_m = 0$. Otherwise, the user will win with utility $\tilde{U}_m(b_m^*, p_m^*)$ and strategy $\tilde{\xi}_m$. This auction scheme maximizes social welfare as it allocates resources to those who value them the most.

The potential *efficiency* of the joint bidding and tipping scheme reflects the probability that winners' demands will be satisfied. This can be characterized from the average probability of accessing the destination, $\eta_{JBiT} = \bar{p}_D$, given by (6). The potential *efficiency* of each strategy is:

$$\eta_{bid} = \int_p \bar{p}_D(p,b)dp \quad (13a)$$

$$\eta_{tip} = \sum_{b'} \bar{p}_D(p,b) \quad (13b)$$

---
**Algorithm 1** *Iterative joint bidding and tipping (i-JBiT)*
---
1. **Initialize:** $t = 1$, $price_c(1) = 0$, $price_q(1) = 0$, $max\_loser = bid_{M+1}$
2. **Store** in $max\_loser$ the highest bid from the user who loses in all previous iterations
3. **Repeat**
4. Each user submits a bid to the auctioneer
5. Let $\{bid_m\}$, $m = 1,…, N$ be the sorted array of bids
6. Find the tentative winner bids from $\{bid_m\}$ / $bid_m \geq price_q$
7. Set the clearing price to $price_q(t) = max\_loser$
8. $t = t + 1$
9. Update the *tentative* clearing price, $price_c(t) = price_q(t-1) + c \cdot t$
10. Each user revises the bid and tip by solving (10) with $\beta_m(t+1)$ and $\theta_m(t+1)$ given by (11) and (12), respectively
11. **Until** $|price_c(t) - price_c(t-1)| < \varepsilon$
12. $price_q = max\_loser$
13. **End**
14. **Return** winners and clearing price $price_q$
---

### C. Economic Properties and Time Complexity

The proofs of the properties truthfulness, individual rationality and computational efficiency are provided in Appendix A.

*Theorem 1*: The *i-JBiT* scheme is truthful such that the SU $m$'s best strategy is $\tilde{\xi}_m = (\tilde{\beta}_m, \tilde{\theta}_m)$.

*Theorem 2*: The *i-JBiT* scheme is individually rational since SU $m$ will not pay more than its true valuation, $bid_m \geq price_q$.

Theorem *3*: The *i-JBiT* scheme is computationally efficient since the auction is solved in polynomial time.

## V. GROUP PARTITIONING-BASED AUCTION DESIGN

The concept of group-buying auctions (e.g., Groupon, Google Offers) is widespread in the context of Internet services [14], [20]. The idea behind online group buying is to recruit enough users to generate a volume of orders large enough to motivate lower transaction prices. These works differ from the group buying schemes applied to spectrum sharing in that there may be more than one item to sell (channel) and that resources can be shared among different users (spectrum reusability). In this section, we define static and dynamic partition schemes to increase SU´s valuation by extending the concept of Internet group buying to spectrum trading in multi-hop networks.

When there is a large number of SU sources bidding for resources at the same time, the PO will reuse the resources to serve the highest bidders. However, high resource reusability implies longer scheduling times, which will degrade SU QoS and its valuation. To deal with this issue, in the sequel we present static and dynamic partition strategies at the PO to group SUs sources and benefit from the heterogeneity of their requirements. Recall that once the winning SUs are determined the multi-hop transmissions are implemented as explained in Section II.C and III.A.

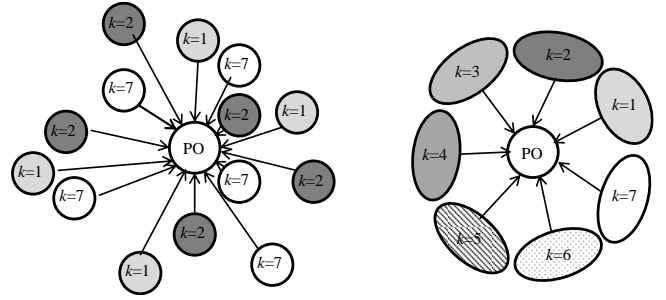

Fig. 3. Static group-bidding scheme based on *K*-reuse pattern.

### A. Static Group Partitioning Scheme

We assume that after the auctioneer receives all bids, it will partition the bidders into $K$ groups, as shown in Fig. 3, following the *K*-reuse pattern introduced in Section II.C (Fig. 1). We call this scheme an *sgroup* scheme. The partition is denoted as $\Lambda_K = \{\mathcal{N}_1, \mathcal{N}_2, ..., \mathcal{N}_k, ..., \mathcal{N}_K\}$, where $k$ is the index of the $k$-th partition and $\mathcal{N}_k$ the set of bidders in that partition. The cardinality of any partition is $|\mathcal{N}_k| = N/K = \lfloor 3R(R+2r)/K \rfloor$, where $N$ is the number of subcells, $R$ the radius of the macrocell and $r$ the radius of the subcells. The users of a winning partition will transmit simultaneously in the same slot $k$ to their adjacent subcells on the way to their respective destinations. This scheme reduces the number of winning groups to $S$, $S \leq K$ and the number of winners to those belonging to the winning groups.

Following the *i-JBiT* scheme, each SU source $m$ chooses its strategy $\xi_m = (\beta_m, \theta_m)$ and determines the number of channels $b_m$ and $p$ to satisfy its QoS by (10). Then, all users simultaneously submit their sealed bids to the auctioneer. After receiving all bids, the auctioneer will apply the previously described static grouping scheme and calculate the bid per group.

We define the bid of a group as the sum of all SU bids in that group. We denote by $\Theta_k$ the bid of group $\mathcal{N}_k$ and by $bid_{m,k}$ the bid of a particular user $m$ within that group. Then, we have

$$\Theta_k = \sum_{m \in \mathcal{N}_k} bid_{m,k} \quad (14)$$

where $bid_{m,k}$ is given by (8) for $m \in \mathcal{N}_k$.

We denote by $\tau_{max,k}$ the maximum delay that the users in group $\mathcal{N}_k$ can tolerate, $\tau_{max,k} = \min\{\tau_{max,m}\}$, where $\tau_{max,m}$ is the QoS requirement of user $m$, $m \in \mathcal{N}_k$.

*Winner Selection:* The auctioneer sorts the group bids in descending order and forms the set $\{\Theta_s\}$, $s = 1,…, K$. Then, the auctioneer selects as potential winners the $S$, $S \leq K$, highest group bids such that $\tau_{max,S} \leq \min\{\tau_{max,s-1}\}$, $s = 1,…, S$ and $m \in \mathcal{N}_s$.

*Payment Mechanism:* The payment mechanism consists of two steps: determination of the price for each winning group and for each SU. The clearing price of a winning group $price_q$ is the highest *group bid* from the *group* that lost in all previous iterations, as given in Algorithm 2. As before, to encourage truthfulness and obtain high revenue, the *tentative* clearing price in each iteration $t$ is sequentially increased as $price_c(t+1) = price_q(t) + \delta \cdot t$.

Given the winning group clearing price defined as above, the price a winning SU $m$ in group $s$ must pay is proportional to its bid,

$$price_{m,s} = \frac{bid_{m,s}}{\Theta_s} price_c \quad (15)$$

*Revision of strategy:* The auctioneer allows *all* SUs to *revise* their strategies $\xi_m(t+1) = (\beta_m(t+1), \theta_m(t+1))$ and submit a new bid and tip as a result of (10). Strategy revision follows *i-JBiT*, where $\beta_m(t+1)$ is now obtained by (11) for $price_{m,s}$, and $\theta_m(t+1)$ by (12).

Once the *i-JBiT* algorithm converges to the clearing price, each source $m$ belonging to a winning group $S$ will transmit to its intended destination on the selected route using the resources purchased. Relays (compensated by tips) will forward the traffic of the winning bidders following the $K$-scheduling pattern. In this scheme, the number of winning groups $S$ is limited by the QoS requirements of the highest bidders. This thus reduces the scheduling time by a factor of $S/K$ and is a fair scheme for users with more restrictive QoS requirements as they will be eager to pay more and this will increase their chances of winning the auction.

---

**Algorithm 2** *Iterative joint bidding and tipping sgroup scheme* (sgroup)

1. **Initialize:** $t = 1$, $price_c(1) = 0$, $price_q(1) = 0$
2. **Store** in $max\_group\_loser$ the highest group bid from the group which lost in all previous iterations
3. **Repeat**
4.     Each user submits a bid to the auctioneer
5.     The auctioneer groups the users following the $K$-reuse pattern and obtains the group bids $\Theta_k$ from (14) and the maximum tolerable delay per group, $\tau_{max,k} = \min\{\tau_{max,m}\}$
6.     Let $\{\Theta_k\}$, $k = 1,\ldots, K$ be the sorted array of group bids
7.     Find the $S$ highest group bids from $\{\Theta_k\} / S \cdot \tau_m \leq \min\{\tau_{max,s}\}$, $s = 1,\ldots, S$
8.     Set the clearing price to $price_q(t) = max\_group\_loser$
9.     $t = t + 1$
10.    Update the *tentative* clearing price, $price_c(t) = price_q(t–1) + c \cdot t$
11.    Obtain the price per user, $price_{m,s}$ by (15)
12.    Each user revises the bid and tip by solving (10) with $\beta_m(t+1)$ and $\theta_m(t+1)$ given by (11) and (12), respectively
13. **Until** $|price_c(t) – price_c(t–1)| < \varepsilon$
14. $price_q = max\_group\_loser$
15. Obtain the price per user, $price_{m,s}$ by (15)
16. **End**
17. **Return** $S$ winner groups and the clearing price per user $price_{m,s}$

---

The PO can create new partitions by changing $r$ and $K$, $(r, K) \rightarrow \Lambda_K = \{\mathcal{N}_1, \mathcal{N}_2, \ldots, \mathcal{N}_K\}$. The effects of these parameters on network performance are the following:

- For constant $K$ and smaller $r$, the number of members per group increases. The delay per route increases with the number of hops and thus, the users will need to pay a higher *tip* as the route will be longer.
- For constant $r$ and changing $K$, the number of hops is fixed. The scheduling duration increases with $K$. The interference and the number of users per group decreases as $K$ increases.

*Theorem* 4: The *i-JBiT* with *sgroup* partition is truthful, individually rational and computationally efficient.

*Proof:* The proof is provided in Appendix B. ∎

### B. Dynamic Group Partitioning Scheme

In this section, we present a dynamic group partitioning scheme (denoted as *dgroup scheme*) that exploits the dynamic arrival of bidders to the network. The auctioneer allows users to join forces within a given time frame $T$ to get volume discounts. This scheme is intended for users with low QoS requirements and a tight budget.

The auction scheme consists of two steps. In the first step, the PO sets its price curve and the auction period $T$. As before, we denote the clearing price in time $t$ as $price_c(t)$. This information is available to the bidders to motivate additional buyers. In the second step, the bidders place bids one by one according to their arrival times. We assume that, when a bidder arrives, it will bid immediately if the resources are considered to be worth the asking price. The bidding and arrival times are therefore the same.

*Winner Selection:* Let us assume that index $m$, $m = 1,\ldots, N$ denotes the index of the SU source and also the sequence number according to its arrival time. To bid successfully, each bidder $m$ must offer a bid $bid_m \geq price_c(t_m)$. The user obtains its strategy from (11) for $\beta_m \geq \beta_{m,c}$. Although the price is given, the optimization in (10) is solved iteratively following the *i-JBiT* to obtain the bid and tip. After this point, the user will wait in the queue until the auction finishes. Otherwise, it will leave the auction immediately and forever. The auction will end after time $T$. When the auction ends, the users whose bids are higher than $price_c(T)$ will be the winners. Note that, as the price decreases in time, a winner in time $t_m$ will also be a final winner.

*Payment Mechanism:* The pricing function is defined as follows. The PO will reuse the resources among the SUs following the $K$-reuse pattern and let the users transmit in the same slot to share expenses. As the number of users increases, the price per user will decrease and, consequently, users will benefit from an increasing number of bidders. The price will also be influenced by the uncertainty of resource availability during the auction period $T$ given by the probability $p_{return}(T)$ [6]. The higher the uncertainty, the lower the price charged for the resources will be. Based on the description above, the average price $price_c(t_m)$ when source $m$ bids at time $t_m$ can be defined as

$$price_c(t_m) = price_c(1) \cdot b_m \cdot \frac{K}{n_S(t_m)} e^{-t_m} \cdot p_{free}(T), \; t_m \leq T \quad (16)$$

where $price_c(1)$ is the initial price of the resources at $t = 1$, $b_m$ is the number of channels that $m$ bids for, $n_S(t_m)$ is the average number of SUs at time $t_m$, and $p_{free}(T) = 1 - p_{return}(T)$.

If we assume that $bid_m \geq price_c(t_m)$ with probability $p_m$, the equivalent rate at which SUs attempt to access the spectrum is

$\lambda_{eq} = \lambda_S \cdot p_m$. Then, the average number of SUs in the network at time $t_m$, $n_S(t_m)$, can be obtained as [19]

$$n_S(t_m) = \sum_{\varsigma=0}^{m} \varsigma p_\varsigma(t_m) = \sum_{\varsigma=0}^{m} \varsigma \frac{(\lambda_{eq} t_m)^\varsigma}{\varsigma!} e^{-\lambda_{eq} t_m} \qquad (17)$$

where $p_\varsigma(t_m)$ is the probability of $\varsigma$ arrivals until time instant $t_m$.

The final price, $price_c(T)$, is obtained when the auction is completed. The winning bidders will transmit to their intended destination and relays (compensated by tips) will forward their traffic following the $K$-scheduling pattern. The average queue waiting time, $w_f$, can be ignored, as $w_f \ll K \cdot \tau_m$.

**Algorithm 3** *Iterative joint bidding and tipping dgroup scheme* (*dgroup*)
1. **Initialize:** number of winners $n_S = 0$
   The auctioneer provides the initial clearing price $price_c(1)$, and duration of the auction $T$
2. **Repeat**
3. A new user $m$ arrives at the network at time $t_m$ and observes the price $price_c(t_m)$
4. Each user revises their bid and tip by solving (10) with $\beta_m(t+1)$ and $\theta_m(t+1)$ given by (11) and (12), respectively.
5. **If** $bid_m \geq price_c(t_m)$ then
6. Update winners by $n_S = n_S + 1$
7. **end**
8. $t_m = t_m + 1$;
9. Update the *tentative* clearing price, $price_c(t_m)$ as in (16)
10. **Until** $t_m < T$
11. Obtain the final clearing price, $price_c(T)$ as in (16)
12. **End**
13. **Return** $n_f$ winners and the clearing price $price_c(T)$

*Theorem* 5: The *i-JBiT* with *dgroup* partition is truthful, individually rational and computationally efficient.

*Proof:* The proof is shown in Appendix C. ∎

Both the static and dynamic partition schemes allow the auctioneer to exploit the heterogeneity of SUs in terms of QoS requirements and willingness to pay in order to increase the operator's revenue and maximize social welfare.

## VI. SIMULATION RESULTS

We conducted Matlab simulations to verify the theoretical analysis, evaluate the performance of our proposed auction schemes and compare them with existing schemes. The network considered is shown in Fig. 1, where the radius of the macrocell is $R = 1000\ m$, $H = 4$ and $P = 0.75$ W. The path loss exponent is $\alpha = 2$ and the noise power is $N_r = 10^{-4}$ W [6]. We assume that in each subcell there is an SU source willing to transmit to the BS or destination user. The density of destinations is set to 0.1. Besides, the source will be available to relay other users' data with probability $p$. We consider that there are $c = 10$ channels in total in the cellular network, $n$ of which are occupied by PUs, and $b$ potentially available channels for SUs. The available channels during a given transmission slot are randomly allocated to the SUs sharing that slot, according to the $K$-reuse pattern. Monte Carlo simulations were run and the results were averaged over 100 iterations.

### A. Joint Bidding and Tipping

The simultaneous impact of $b$ and $p$ on average delay $\bar{\tau}$ is shown in Fig. 4. Note that a QoS requirement given by $\tau_{max}$ can be achieved by different combinations of $p$ and $b$. In particular, the higher $b$ the lower $p$ is needed to achieve the same delay, and vice versa. The minimum and maximum delay per user observed in this network was 1 and 7, respectively. This gap will be reflected in the QoS requirements.

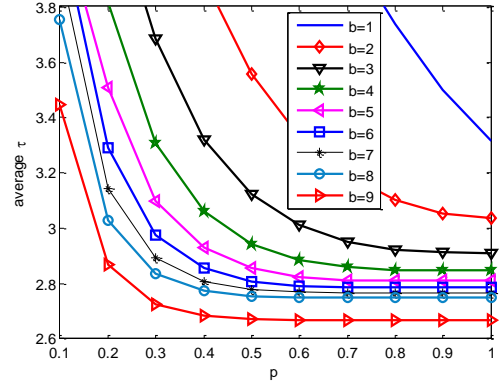

Fig. 4. Average $\tau$ versus $p$ for different values of $b$ where $\tau$ defined by (4).

In the following, we assume $K = 7$ and users' QoS requirement $\tau_{max}$ randomly varying between [7, 49]. The joint bidding and tipping scheme is run for $T = 25$ iterations until the trading price is reached, $|price_c(t) - price_c(t-1)| < \varepsilon$. The price step is $\delta = 10^{-3}$ and the initial value of $\beta_m$ and $\theta_m$ is set to 0.05. The revision of the bid and tip in each iteration is shown in Fig. 5, together with the tentative clearing price. We can see that the bid increases in each iteration to follow the tentative clearing price. To keep raising the bid, the user will moderate the number of channels it bids for, as shown in Fig. 6. In addition, to continue to meet the QoS requirement, the user will request a higher relay availability probability $p$. As a result, the tip will iteratively decrease, as shown in Fig. 5. Note that the percentage of the valuation offered as a tip is $\theta = 1 - p$.

In Fig. 6, the optimum number of channels $b^*$ is shown for different $p^*$ as a result of solving (10) by *i-JBiT*. We can see that for higher $p^*$, fewer channels $b^*$ are needed to satisfy the QoS requirement, as previously discussed. The average user utility is shown in Fig. 7 versus the iteration index for different relay availability probabilities $p$. By increasing $p$, a higher utility is obtained as the user's valuation increases. Since the *i-JBiT* algorithm iteratively updates the bid and tip, the utility oscillates during the process.

In Fig. 8, the revenue of the auctioneer is presented versus the iteration index. The revenue increases with the iteration index and $p$. However, for $p > 0.7$, the gain is not significant as the user needs fewer channels.

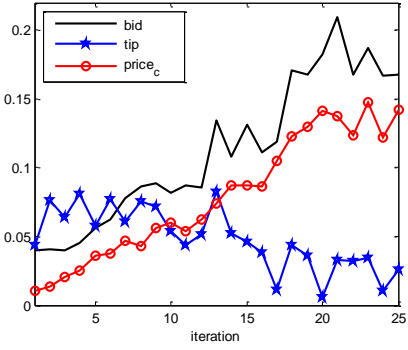
Fig. 5. Bid, tip and price vs. iteration when $p = 0.7$.

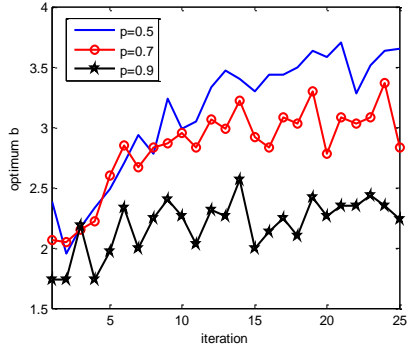
Fig. 6. Optimum $b$ for different $p$ vs. iteration.

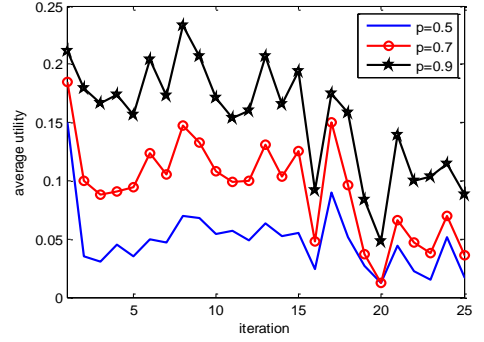
Fig. 7. Average utility versus iteration.

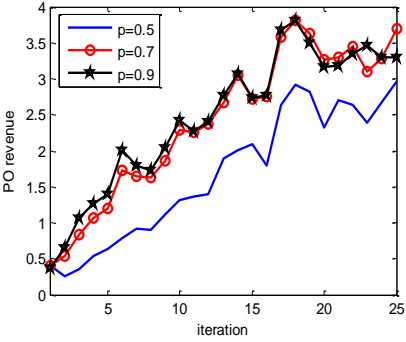
Fig. 8. PO revenue vs. iteration.

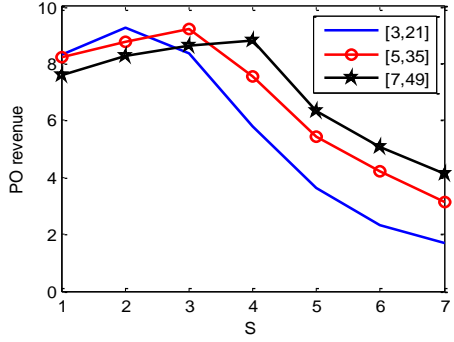
Fig. 9. PO revenue in *sgroup* scheme vs. $S$.

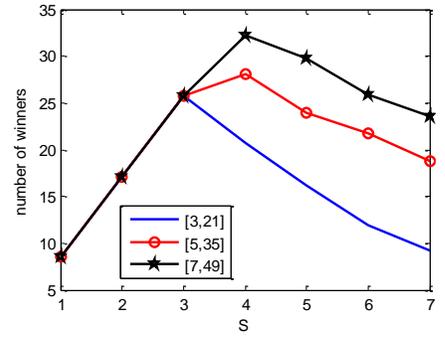
Fig. 10. Number of winners in *sgroup* scheme vs. $S$.

### B. Group Partitioning Schemes

We consider three scenarios to show the performance of the auction when the PO utilizes the *sgroup* partition scheme. In each scenario, we assume that the QoS requirements of the SUs $\tau_{max}$ vary randomly in [3, 21], [5, 35] and [7, 49], respectively.

The revenue of the PO is shown in Fig. 9 for the previous scenarios versus the number of winning groups, $S$. The highest revenue is obtained for the first scenario when $S = 2$. This scenario is the one with the most restrictive QoS requirements. As the QoS requirements become more relaxed (scenario 1 → 3), the optimum $S$ increases since users can tolerate higher delays. We can also see that the PO revenue is about 3 times higher than in the scheme without groups ($S = 7$) in the most restrictive case. In scenario 3, even though the users can tolerate higher delays, the PO revenue is twice as high as in the case without grouping. The number of winners in each case is shown in Fig. 10. In the first scenario, the *sgroup* scheme provides 150% more winners than the no-group scheme and 40% more than in scenario 3. It is worth noting that although the number of winners in scenario 1 is the lowest, the PO obtains the highest revenue as the users place high value on the resources. To show the performance of the *dgroup* partition scheme, we considered three additional scenarios. As this scheme is intended to increase the revenue of the PO when users have low QoS requirements, in the new scenarios we assume that the QoS requirements $\tau_{max}$ vary randomly between [9,63], [11,77] and [13,91]. The final clearing price $price_c(T)$ is set to a fraction of the final clearing price obtained in the scheme without groups for a fair comparison. In Fig. 11, the number of winners is shown for the different scenarios. We can see that by reducing the price per channel by 50%, the number of winners increases by up to 45% in the low demand scenarios, with consequent increase in social welfare. The revenue obtained by the PO is shown in Fig. 12 for the same scenarios. When the clearing price is reduced by 25%, the increased number of winners compensates the price reduction and the revenue obtained is even higher than in the scheme without groups.

The initial values of $\beta_m$ and $\theta_m$ have impact on the convergence speed of the bidding and tipping process. The PO can adjust the speed by changing the price step $\delta$. In order to keep the utility $U_m \geq 0$, we need $\beta_m = \theta_m \leq 1/(b_m + \tau_m)$. For the range of QoS requirements considered, $\beta_m = \theta_m \leq 0.25$ when $\tau_m = 3$ and $b = 1$, and $\beta_m = \theta_m \leq 0.01$ when $\tau_m = 91$ and $b = 9$. Our simulations suggest that if the initial value of $\beta_m$ and $\theta_m$ is set to 0.25, similar convergence speed as in the current case with 0.05 is achieved. However, if we change the initial value of $\beta_m$ and $\theta_m$ to 0.01 similar convergence speed can be obtained decreasing $\delta$ to 1/10 times of its previous value.

### C. Comparison with other schemes

To the best of our knowledge, our auction schemes are the first to consider bidding and tipping strategies as well as PO partition strategies to exploit heterogeneous QoS requirements in multi-hop secondary networks. As the previous section already provides insights into PO gains obtained with and without partition schemes, here we focus on comparing our model, which integrates route discovery and the *i-JBiT* algorithm, with other approaches.

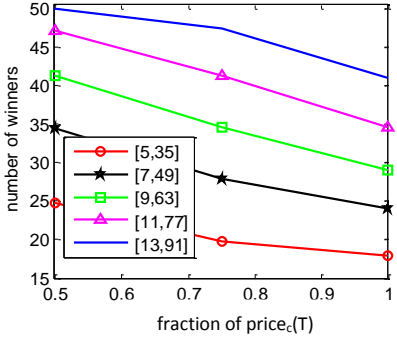
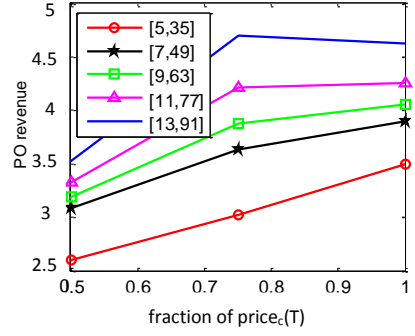
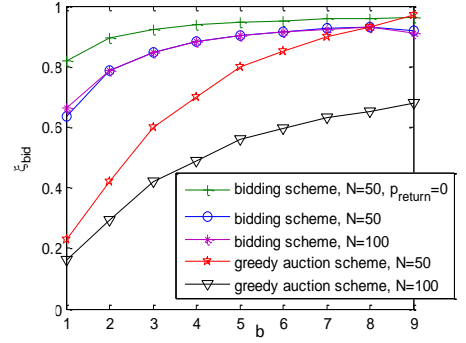

Fig. 11. Number of winners in *agroup* scheme.   Fig. 12. PO revenue in *dgroup* scheme.   Fig. 13. Bidding efficiency comparison.

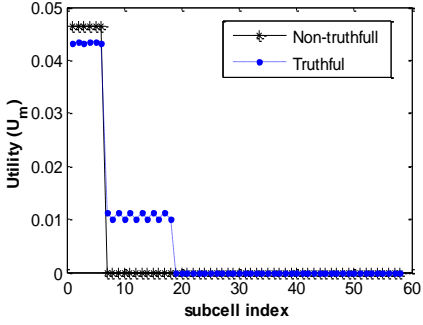
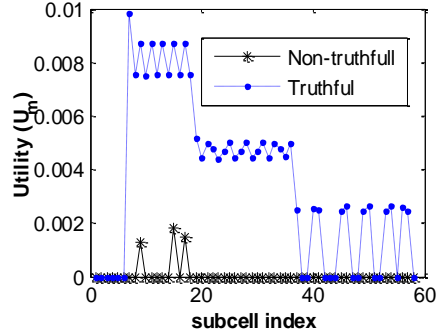
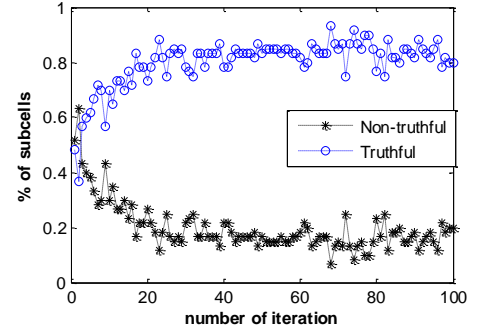

Fig. 14. $U_m$ versus subcell index $m$.   Fig. 15. $U_m$ versus subcell index $m$.   Fig. 16. Percentage of truthful and untruthful bidders.

Figure 13 shows the potential efficiency of the proposed bidding scheme compared with the scheme in [13], which greedily assigns channels to different links. Recall that in our bidding scheme users estimate the resources needed and select the best route thanks to the route discovery protocol. We can see that our scheme significantly outperforms the greedy scheme as it considers the existence of multiple possible routes to the destination depending on the availability probability of the relays. Besides, we can observe the robustness of our scheme as the network grows, while the efficiency of the greedy scheme drops for large user populations. By applying a *K*-scheduling pattern, we keep interference under control. The highest interference levels come from the first tier of interfering users as shown in Fig. 1, and therefore increasing network size does not significantly increase overall interference. We can also see that, for $b=1$, efficiency increases slightly with $N$. This is because we are increasing the length of the relaying routes, which is especially critical when there is just one channel available. For a fair comparison, we also show the performance of our scheme when there is no PU return, as considered in the greedy scheme. As expected, an additional increase in efficiency is also seen.

Next, we compare our *i-JBiT* algorithm, which has been proved to be truthful, with a reinforcement learning-based bidding algorithm. There are a number of works that consider reinforcement learning (RL) for spectrum allocation [21], [22]. In our paper, RL is used to develop a bidding algorithm that enables the SUs to revise their bids (and truthful/untruthful behavior) based on previous experience.

We establish two different fixed values of $\beta$: one for the truthful bid ($\beta_t$) and another smaller one for the untruthful bid ($\beta_u < \beta_t$). Therefore, in an attempt to improve its utility, an SU must decide which action to take, since there is a tradeoff between obtaining resources (using a higher $\beta$) and increasing utility (using a smaller $\beta$). For this purpose, we endow SUs with learning capabilities. The value of $\theta$ is updated as in (12). We denote the probabilities that the SU source $m$ will bid truthfully or untruthfully as $p_{m,t}$ and $p_{m,u}$, respectively. Each SU updates both probabilities individually using the Learning Automata algorithm [23]. For example, suppose that a user obtained a higher utility by using $\beta_u$, then the probabilities will be updated as

$$p_{m,u}(t+1) = p_{m,u}(t) + \delta \cdot (1 - p_{m,u}(t)) \tag{18a}$$

$$p_{m,t}(t+1) = p_{m,t}(t) \cdot (1 - \delta) \tag{18b}$$

where $0 < \delta < 1$ is a step size parameter. Algorithm 4 describes the bidding process. SU agents update their actions following an iterative process, which involves bidding and learning.

---

**Algorithm 4** *Reinforcement learning-based auction.*

1. **Input**: $\beta$
2. **for** $j = 1$ : Number_of_Iterations do
3.    $[U, a]$=bidding($\beta$)
4.    **for** $m = 1 : N$
5.      **if** used_Both_Actions() **then**
6.        $[p_{m,t}, p_{m,u}]$ = update_Learning_Probabilities($a_m$, $U_m$)
7.        $\beta_m$ = choose_Action($p_{m,t}, p_{m,u}$)
8.      **end**
9.    **end**
10. **end**

---

We consider two scenarios in order to study how learning affects the bidding process and we also study the process from a user perspective. In the first scenario, the QoS requirements of the users in ring 1, $\tau_{max,1}$, are more restrictive than those for the users in rings 2, 3 and 4 (i.e., $\tau_{max,4} > \tau_{max,3} > \tau_{max,2} > \tau_{max,1}$). To achieve this QoS, we consider that all users bid for the same number of channels ($b_m = 3$). As result, users in ring 1 place a much higher value on the resources than those in the other rings. We set the truthful and untruthful cost percentages to $\beta_t = 0.04$ and $\beta_t > \beta_u = 0.02$, respectively. Note that the value of $\beta_u$ was chosen to be significantly lower than $\beta_t$ to show the effects of untruthful bidding. The number of iterations needed for the learning process to converge is less than 100.

Figure 14 shows the utility per user $U_m$ after the learning process. Subcell indexes 1-6 correspond to the users from the first ring, 7-18 to the users from the second ring, 19-36 to the users from the third ring and the rest to the users from the fourth ring. The simulations indicate that in this scenario a relatively small fraction of the users (approximately 1/3) can afford the resources. These users are located in the first and second rings around the BS. We observed that users in the first ring learn that using $\beta_u$ is the best option. This is because, by using $\beta_u$ rather than $\beta_t$, they increase their utility and still bid high enough to obtain the resources. In the second ring, their valuation of resources is lower, so they learn that, in order to improve their utility, they need to bid using $\beta_t$. By using a smaller value, they will not obtain the resources and, therefore, their utility will be zero. Finally, for the last two rings, we see that, regardless of the percentage $\beta$ chosen, users will always obtain $U_m = 0$. This is because their valuation of resources is insufficient to win. Thus, in scenarios with misbalanced competition, a reinforcement learning-based bidding algorithm cannot guarantee truthfulness.

In the second scenario, the competition among users is more balanced. We assume that users require similar QoS and thus, users from different rings will demand different numbers of channels. In particular, we set the number of channels to $b_m = 1$ for users in the first ring, $b_m = 2$ for users in the second ring, $b_m = 3$ for users in the third ring and, finally, $b_m = 7$ for users in the last ring. The truthful and untruthful cost percentages were set to $\beta_t = 0.02$ and $\beta_u = 0.01$, respectively. Figure 15 shows the utility per user $U_m$ after the learning process. We can see that, in this case, the users in the first ring do not value the resources highly enough to be able to compete for them. However, the users in the other rings learn that they can increase their utility by using $\beta_t$. In Fig. 16, we show how the percentages of truthful and untruthful bidders evolve through iterations of the learning process. As this scenario allows fairer competition among rings, the percentage of users that bid truthfully increases. In a real network, users will have heterogeneous valuations of resources, so auction mechanisms such as *i-JBiT* should be applied to ensure truthfulness.

## VII. RELATED WORK

Recently, multi-hop cellular networks (MC$^2$N) have attracted a lot of attention due to their potential to achieve efficient spectrum usage by exploiting local available channels and support dynamic traffic distributions without additional infrastructure costs [24], [25]. Most auction applications in MC$^2$N are incorporated for routing rather than spectrum allocation purposes [26]-[27]. To exploit the full potential of MC$^2$N, bidding strategies for spectrum allocation, route selection and relaying incentives should be jointly considered.

Designing auction mechanisms for MC$^2$N that consider QoS is a challenging task as the process involves bidding for multiple channels. The few works which consider spectrum auction in multi-hop networks [12], [13] assume that users bid for a single channel, and ignore SU QoS requirements. Heterogeneous demands in spectrum auctions have been considered for single-hop networks in [15]-[16] and [28]-[29]. In [28] a QoS-aware auction framework is developed for bidders to dynamically bid for primary or secondary users rights according to their QoS demands. A multi-channel auction scheme is presented in [29] to satisfy heterogeneous SU demands in terms of number of requested channels. The auction scheme is based on a combinatorial auction which is computationally complex and channel reuse is not considered. In [15], [16], an auction framework for spectrum group buying is presented where buyers are voluntarily grouped together to acquire and share the spectrum band sold in the auction. However, even though in this case buyers bid for multiple channels, they will only win one channel and the main purpose of these schemes is to obtain resources at a lower price. Our partition-based auction schemes, in addition to providing group discounts, allow the partitioning of users based on their QoS requirements, thereby ensuring a fair distribution of resources. Despite being multi-item auctions, our schemes are truthful, incentive compatible and computationally efficient. This last property is highly desirable for auctions in multi-hop networks.

## VIII. CONCLUSIONS

This paper presents a new concept of spectrum auction in multi-hop cognitive cellular networks (MC$^2$Ns). First, a joint bidding and tipping scheme is developed to iteratively revise bidding and tipping strategies. Based on the clearing price set by the auctioneer and the incentive needed to encourage the relays to cooperate, SUs decide the number of resources needed to satisfy their QoS requirements in the most profitable way. Then, group partitioning-based auction schemes are presented together with new winner selection and payment mechanisms to exploit the heterogeneous QoS demands of SUs.

An extensive set of simulation results is provided to evaluate different schemes. Simulation results have shown that for highly demanding users, compared with the no-group scheme, the static group scheme provides 3 times higher revenue for the PO and 150% more winners with a consequent increase in social welfare. For low-demanding users, the PO can keep a similar revenue with the dynamic scheme by lowering the price per channel by 50% as the number of winners will increase proportionally. We also show that our *i-JBiT* scheme outperforms other schemes for strategy revision based on reinforcement learning and is truthful, incentive compatible and computationally efficient. As a byproduct, our model, which integrates a route discovery protocol, allows for tractable analysis and is more robust than other models when large populations of bidders are considered.

APPENDIX A: PROOF OF PROPERTIES FOR *i-JBiT* SCHEME

*1) Proof of Theorem 1 (Truthfulness)*: This proof follows from [13] and [31]. Nevertheless, we decided to include it for sake of completeness. To prove the truthfulness of the *i-JBiT* scheme we first show that the resources are monotonically allocated and winners are charged with critical value.

*Definition 1:* **Monotonic allocation:** When the bidding strategies of other users $\boldsymbol{\beta}_{-m}$ are fixed, if user *m* wins the auction by bidding $bid_m(\tilde{\beta}_m)$, then it will also win by bidding $bid_m(\beta_m') > bid_m(\tilde{\beta}_m)$.

*Definition 2:* **Critical value:** The critical value is such that if the user bids higher than that, it wins and otherwise, it loses.

*Lemma 1:* The auction resources, i.e., channels, are monotonically allocated in our *i-JBiT* scheme.

*Proof:* The *i-JBiT* scheme determines that $bid_m(\tilde{\beta}_m)$ is a winning bid when $bid_m(\tilde{\beta}_m) \geq price_q$. Thus, if $bid_m(\beta_m') > bid_m(\tilde{\beta}_m)$ then $bid_m(\beta_m')$ is also a winning bid.

*Lemma 2:* The clearing price of the *i-JBiT* scheme, $price_q$, is a critical value.

*Proof:* Since user *m* wins the auction when $bid_m(\tilde{\beta}_m) \geq price_q$, this lemma directly follows. Besides, if user *m* wins the auction by bidding $bid_m(\tilde{\beta}_m)$ and $bid_m(\beta_m')$, then the payment $price_q$ is the same for both.

Using the previous claims, let us prove the truthfulness of *i-JBiT*. We prove that if the best bidding strategy of user *m* is $\tilde{\beta}_m$, following a strategy $\beta_m'$, $\beta_m' \neq \tilde{\beta}_m$ results in $U_m(\beta_m') \leq U_m(\tilde{\beta}_m)$. The following cases are possible:

- If user *m* places a bid $bid_m(\beta_m') < bid_m(\tilde{\beta}_m)$ with $bid_m(\beta_m') > price_q$, then according to Lemma 1 and 2, *m* is charged the same price and thus, $U_m(\beta_m') = U_m(\tilde{\beta}_m)$. Our claim holds.

- If user *m* places a bid $bid_m(\beta_m') < bid_m(\tilde{\beta}_m)$ with $bid_m(\beta_m') < price_q$, then according to Lemma 1 and 2, *m* will loss and its utility is zero ($U_m(\beta_m') = 0$). Our claim holds.

- If user *m* places a bid $bid_m(\beta_m') > bid_m(\tilde{\beta}_m)$ with $bid_m(\beta_m') > price_q$, then according to Lemma 1 and 2, *m* is charged the same price and thus, $U_m(\beta_m') = U_m(\tilde{\beta}_m)$. Our claim holds.

- If user *m* places a bid $bid_m(\beta_m') > bid_m(\tilde{\beta}_m)$ with $bid_m(\beta_m') < price_q$, then $\tilde{\beta}_m$ cannot be the best bidding strategy of user *m* and thus, this case is not possible.

Hence, a user cannot improve its utility by submitting a bid different from its true bid and we can conclude that *i-JBiT* is truthful. ∎

*2) Proof of Theorem 2 (Individual rationality)*: Since the bids are sorted in decreasing order, $bid_m(\tilde{\beta}_m) \geq bid_{m+1}(\tilde{\beta}_{m+1})$ and the price is determined as $price_q = bid_l$, where $bid_l$ is the highest loosing bid, $bid_l < bid_{m+1}$ then, $price_q < bid_{m+1} \leq bid_m$. We can conclude that *i-JBiT* is individually rational since any user *k* will not pay more than its bid. ∎

*3) Proof of Theorem 3 (Computational efficiency)*: We analyze the running time of *i-JBiT*. Winner selection (identifying and ordering highest bids) takes $O(\log N)$ steps. In the worst case, the process is repeated *T* times and then, *M* winners are selected, which takes $O(TM \log N)$ steps. When the users revise their strategies, they need to solve problem (10), which is a convex problem and can be solved in polynomial time [30]. Consequently, the overhead is acceptable.

APPENDIX B: PROOF OF PROPERTIES OF *i-JBiT* SGROUP SCHEME (THEOREM 4)

*1) Proof of Truthfulness:* The proof of truthfulness of the *i-JBiT sgroup* scheme can be easily derived from the proof for the *i-JBiT* scheme considering a strategy per group $\tilde{\beta}_s = \{\tilde{\beta}_1, \tilde{\beta}_2, ..., \tilde{\beta}_{ns}\}$ and assuming that one of the users (e.g., user *m* = 1) chooses a strategy, $\beta_1' \neq \tilde{\beta}_1$, which results in strategy $\beta_s' = \{\beta_1', \tilde{\beta}_2, ..., \tilde{\beta}_{ns}\}$. The same reasoning applies for any other user or for situations in which several users simultaneously change strategies. The winning strategy $\tilde{\beta}_s$ is such that $\Theta_s(\tilde{\beta}_s) \geq price_q$. The details are omitted due to space constraints. ∎

*2) Proof of Individual rationality:* Let us denote by $\Theta_s$ a *winning group* bid obtained as $\Theta_s = \sum_{m \in \mathcal{N}_s} bid_{m,s}$. According to the winner selection and payment scheme, the clearing price is $price_q = \Theta_l \leq \Theta_s$ $\forall s$, where $\Theta_l$ is the highest losing group bid. The price of each user *m* in the winning group *s* is given by $price_{m,s} = \frac{bid_{m,s}}{\Theta_s} price_q$, and thus $price_{m,s} \leq bid_{m,s}$. We can conclude that the *i-JBiT sgroup scheme* is individually rational since any user *m* will not pay more than its bid. ∎

*3) Computational efficiency:* We analyze the running time of the *i-JBiT* sgroup. Winner selection (grouping and ordering) takes $O(NK \log S)$ steps. In the worst case, the process is repeated *T* times which takes $O(TNK \log S)$ steps. When the users revise their bids, they need to solve problem (10), which is a convex problem and can be solved in polynomial time [30]. Finally, if there are *ns* winning users per group *S*, to obtain the bid per user renders an overall complexity of $O(TNK \log S + Sns)$. Consequently, although the grouping adds additional complexity the overhead is still acceptable.

APPENDIX C: PROOF OF PROPERTIES FOR DGROUP SCHEME (THEOREM 5)

*1) Proof of Truthfulness:* Following the price set up of the *dgroup* scheme given in (16), if $\tilde{\beta}_m$ is the winning bidding strategy, then $bid_m(\tilde{\beta}_m) \geq price_c(t_m)$. If the user bids with another strategy $\beta_m' \neq \tilde{\beta}_m$ such that, $bid_m(\beta_m') < price_c(t_m)$, then $U(\beta_m') = 0$. Thus, a user cannot improve its utility by placing other than its true bid and we can conclude that *dgroup* is truthful. ∎

*2) Proof of individual rationality:* The *dgroup* scheme uses a decreasing pricing mechanism. If the user bids successfully at time $t_m$, $bid_m(\tilde{\beta}_m) \geq price_c(t_m)$, it will also win at time $T$ as the price decreases in time and with the number of bidders (16). Hence, $bid_m(\tilde{\beta}_m) \geq price_c(T)$ and we can conclude that *dgroup* is individually rational since any user *m* will not pay more than its bid. ∎

*3) Computational efficiency:* The computational complexity of the *dgroup* scheme is the lowest of the three schemes. The winner and payment calculation takes $O(n_f \cdot N)$ steps. The user bid is obtained by solving problem (10), which is a convex problem and can be solved in polynomial time [30]. Consequently, the overhead is acceptable.